\documentclass[aps,prl,twocolumn,superscriptaddress]{revtex4}%
\usepackage{graphicx}
\usepackage{dcolumn}
\usepackage{amsmath}
\usepackage{amssymb}
\usepackage{indentfirst}
\usepackage{bm}
\usepackage{multirow}
\usepackage{graphicx}
\usepackage{hyperref}
\hypersetup{hypertex=true,
colorlinks=true,
linkcolor=blue,
anchorcolor=blue,
citecolor=blue}

\begin{document}

\title{Nonadiabatic Holonomic Single-Qubit Gates in Non-Hermitian Systems\footnote{This is a preliminary version. Please do not cite without permission.}}

\author{Wei Li} \email{liweixici@126.com}
\affiliation{
School of Physics and Electronic Information, Yanan University, Yanan 716000, China}

\author{Yue Zhang}
\affiliation{
School of Physics and Electronic Information, Yanan University, Yanan 716000, China}

\author{Yu Kun Chen}
\affiliation{
School of Physics and Electronic Information, Yanan University, Yanan 716000, China}

\author{Jia Yao Liang}
\affiliation{
School of Physics and Electronic Information, Yanan University, Yanan 716000, China}

\date{\today}

\begin{abstract}
Holonomic quantum computation offers a promising route to robust quantum gates, but decoherence remains a central obstacle in realistic implementations. Here we develop a nonadiabatic holonomic scheme for a driven three-level system in the no-jump regime described by an effective non-Hermitian Hamiltonian. Within a biorthogonal framework, tailored complex pulses enforce exact closure of the computational-subspace evolution at the final time despite the underlying nonunitary dynamics, enabling arbitrary single-qubit holonomic gates without requiring cyclic evolution in its orthogonal complement. In contrast to existing non-Hermitian treatments, which either neglect the overall exponential prefactor or, in adiabatic settings, include dissipation only on the auxiliary excited level, our scheme incorporates decay and dephasing of all bare eigenstates directly into the pulse design, so that dissipation does not reduce the no-jump gate fidelity.
\end{abstract}
\maketitle
\section{INTRODUCTION}
Quantum computation can outperform classical computation on tasks such as integer factorization \cite{integers}, unstructured search \cite{data}, certain sampling problems \cite{Complexity}, and quantum simulation \cite{Simulating}. Achieving such advantages requires gate operations that are both fast and robust against control imperfections and decoherence. Geometric quantum gates address this challenge by encoding operations in phases determined by global features of the evolution path, thereby providing intrinsic resilience to certain classes of errors \cite{local1,local2,local3}.

Geometric quantum computation was originally formulated in the adiabatic regime \cite{GQC1,GQC2,GQC3,GQC4}. Cyclic adiabatic evolution of a nondegenerate eigenstate gives rise to the Berry phase \cite{Berry}, while its non-Abelian generalization in a degenerate subspace \cite{matrix} underlies adiabatic holonomic quantum computation. The long operation times imposed by adiabaticity, however, increase exposure to decoherence and have motivated nonadiabatic geometric schemes \cite{AA,single1,single2,single3,single4,single5,multiple1,multiple2,multiple3,holonomic1,holonomic2,composite,NHQC,NHQC+,schemes1,schemes2,schemes3}, which preserve the geometric character of the control while enabling substantially faster gate operations. Geometric gates have now been demonstrated across a wide range of platforms, including nuclear magnetic resonance \cite{resonance1,resonance2,resonance3,resonance4}, trapped ions \cite{ions1,ions2,ions3}, superconducting circuits \cite{circuits1,circuits2,circuits3,circuits4,circuits5}, solid-state spins in diamond \cite{diamond1,diamond2,diamond3,diamond4,diamond5}, semiconductor quantum dots \cite{dots}, and neutral-atom or optical-lattice systems \cite{atoms1,atoms2}.

Extending geometric control beyond Hermitian systems is both conceptually compelling and practically unavoidable. Realistic driven systems inevitably involve population loss and phase decoherence, while conditioned no-jump evolution is effectively non-Hermitian. Within the biorthogonal formalism, Abelian geometric phases and non-Abelian holonomies admit a unified description in both non-Hermitian and open systems \cite{open1,open2,open3,open4,open5}. Yet existing non-Hermitian geometric-gate schemes face two main limitations. In effective dissipative two-level systems, nonadiabatic geometric gates can be realized, but the no-jump dynamics generally comprises not only an $SU(2)$ sector but also an unavoidable scalar attenuation factor, rendering the evolution globally nonunitary and potentially limiting gate performance \cite{Hou}. Beyond this minimal setting, the existing holonomic proposal is restricted to an adiabatic four-level tripod configuration and does not incorporate decay and dephasing in the computational states \cite{adiabatic}. It therefore remains unclear whether holonomic control can be extended to a genuinely nonadiabatic three-level setting that overcomes loss-induced limitations while directly accommodating dissipation in both computational and auxiliary states.

In this Letter, we show that a non-Hermitian three-level system already suffices to implement arbitrary nonadiabatic holonomic single-qubit gates in the no-jump regime. Within a biorthogonal framework, we design tailored complex control pulses that enforce exact closure of the computational-subspace evolution at the final time despite the underlying nonunitary dynamics. We further incorporate decay and dephasing of both the computational and auxiliary states directly and systematically into the pulse design, so that these dissipative processes need not reduce the no-jump gate fidelity. Conceptually, this construction generalizes the standard instantaneous holonomic constraint \cite{holonomic1} to a non-Hermitian three-level setting, while avoiding the piecewise dynamical-phase cancellation commonly used in NHQC+ schemes \cite{NHQC+}. Using a physically motivated Rydberg-atom implementation, we further show that the resulting dissipation-control matching yields robust performance against representative systematic errors and high fidelities under full quantum-jump dynamics.

\textit{General Framework}. Let us consider a non-Hermitian quantum system governed by a time-dependent Hamiltonian $H(t)$ in an $N$-dimensional Hilbert space $\mathcal{H}$, decomposed as $\mathcal{H}=\mathcal{M}(t)\bigoplus\mathcal{C}(t)$, where $\mathcal{M}(t)$ is an $M$-dimensional subspace and $\mathcal{C}(t)$ its complement. Holonomic gates require a cyclic evolution of the computational subspace, $\mathcal{M}(t_f)=\mathcal{M}(t_0)$. We introduce a biorthogonal basis $\{|\psi_{n}(t)\rangle, |\tilde{\psi}_{n}(t)\rangle\}$
spanning $\mathcal{M}(t)$, satisfying
\begin{equation}
\langle\tilde{\psi}_{m}(t)|\psi_{n}(t)\rangle=\delta_{mn}, \label{biorthonormal}
\end{equation}
with $|\psi_{n}(t)\rangle$ and $|\tilde{\psi}_{n}(t)\rangle$ obeying the Schr\"{o}dinger equation $i\partial_{t}|\psi_{n}(t)\rangle=H(t)|\psi_{n}(t)\rangle$ and its adjoint $i\partial_{t}|\tilde{\psi}_{n}(t)\rangle=H^{\dag}(t)|\tilde{\psi}_{n}(t)\rangle$, respectively. The projected evolution operator on $\mathcal M(t)$ is then
$U_s(t,t_0)=\sum_n |\psi_n(t)\rangle\langle \tilde\psi_n(t_0)|$.

To extract the geometric structure of the subspace evolution, we introduce an auxiliary biorthogonal frame $\{|\nu_{n}(t)\rangle, |\tilde{\nu}_{n}(t)\rangle\}$ satisfying the cyclic boundary conditions
\begin{eqnarray}
\begin{aligned}
|\psi_{n}(t_0)\rangle&=|\nu_{n}(t_0)\rangle=|\nu_{n}(t_f)\rangle,  \\
|\tilde{\psi}_{n}(t_0)\rangle&=|\tilde{\nu}_{n}(t_0)\rangle=|\tilde{\nu}_{n}(t_f)\rangle. \label{boundary}
\end{aligned}
\end{eqnarray}
Expansion $|\psi_{n}(t)\rangle=\sum_{k} W_{kn}(t)|\nu_{k}(t)\rangle$ and substituting it into the Schr\"{o}dinger equation, we obtain
\begin{equation}
W(t)=\mathcal{T}\exp\left[i\int_{t_{0}}^{t}(\tilde{A}(t')-\tilde{H}(t'))dt'\right]. \label{W}
\end{equation}
where $\tilde A(t)$ and $\tilde H(t)$ are the geometric connection and the projected Hamiltonian in the auxiliary frame, with matrix elements $\tilde{A}_{kn}(t)=i\langle\tilde{\nu}_{k}(t)|\dot{\nu}_{n}(t)\rangle$ and $\tilde{H}_{kn}(t)=\langle\tilde{\nu}_{k}(t)|H(t)|\nu_{n}(t)\rangle$, respectively.

In general, both $\tilde A(t)$ and $\tilde H(t)$ are non-Hermitian, so the projected dynamics is nonunitary at intermediate times. Here a purely geometric gate does not require the instantaneous generator $\tilde A(t)-\tilde H(t)$ to coincide with the Hermitian part of the geometric connection, $\tilde A_{\mathrm H}(t)=[\tilde A(t)+\tilde A^\dagger(t)]/2$, at every instant. It is sufficient that the corresponding time-ordered propagator coincide with that generated by $\tilde A_{\mathrm H}(t)$ at the end of the cycle. To formulate this condition, we further introduce a biorthogonal basis $\{\nu^{c}_{n}(t)\rangle, |\tilde{\nu}^{c}_{n}(t)\rangle\}$ of $\mathcal{C}(t)$, and define
\begin{eqnarray}
V(t)=\sum_{n}|\nu_{n}(t)\rangle\langle n|+ \sum_{m}|\nu^{c}_{m}(t)\rangle\langle m^{c}|, \nonumber \\
\tilde{V}(t)=\sum_{n}|n\rangle\langle\tilde{\nu}_{n}(t)|+ \sum_{m}|m^{c}\rangle\langle\tilde{\nu}^{c}_{m}(t)|,  \label{operators}
\end{eqnarray}
where $\{|{n}\rangle\}$ and $\{|m^{c}\rangle\}$ are fixed orthonormal bases of $\mathcal{M}(t_0)$ and $\mathcal{C}(t_0)$. These operators define an effective Hamiltonian in the auxiliary frame,
\begin{equation}
H^V(t)=i\tilde V(t)\dot V(t)-\tilde V(t)H(t)V(t),
\end{equation}
such that $(\tilde A(t)-\tilde H(t))_{kn}=\langle k|H^V(t)|n\rangle$. We then impose the final-time condition
\begin{equation}
\mathcal{T}\exp\!\left[i\int_{t_0}^{t_f}\tilde A_{\mathrm H}(t)\,dt\right]
=
\mathcal{T}\exp\!\left[i\int_{t_0}^{t_f}H^V(t)\,dt\right],
\label{finalcondition}
\end{equation}
which guarantees that the implemented gate is purely holonomic.

In this work, we focus on diagonal holonomies, for which $W(t_f)=\mathrm{diag}(e^{i\Phi_1},e^{i\Phi_2}\ldots e^{i\Phi_M})$ with $\Phi_l=\int_{t_0}^{t_f}\mathrm{Re}[\,i\langle\tilde\nu_l(t)|\dot\nu_l(t)\rangle\,]dt$, and thus take the effective Hamiltonian $H^V(t)$ to be diagonal, with matrix elements $H^V_{mm}(t)=f_m(t)$. The holonomic condition then reduces to $\int_{t_0}^{t_f}f_m(t)dt=\Phi_m$. This integral constraint still leaves considerable freedom in the choice of $f(t)$, allowing the intermediate evolution to be nonunitary while providing flexibility in Hamiltonian design, for instance to enhance robustness against systematic errors. Although the evolution is generally nonunitary at intermediate times, the gate acting on the computational subspace is unitary at the end of the cycle provided that the initial auxiliary frame is orthonormal, namely
\begin{eqnarray}
\begin{aligned}
|\tilde{\nu}_{k}(t_0)\rangle&=|\nu_{k}(t_0)\rangle, &
\sum_{k}|\nu_{k}(t_0)\rangle \langle\nu_{k}(t_0)|&=I_{\mathcal{M}}.  \label{boundary1}
\end{aligned}
\end{eqnarray}
The resulting evolution operator is therefore
\begin{equation}
U_s(t_f,t_0)=\sum_k e^{i\Phi_k}\,|\nu_k(t_0)\rangle\langle\nu_k(t_0)|. \label{unitary1}
\end{equation}
Note that, although Eq. (\ref{unitary1}) is diagonal in the initial auxiliary basis, the holonomies generated by different closed auxiliary frames are generally noncommuting in a fixed logical basis, manifesting their non-Abelian nature.

In general, for a given physical Hamiltonian $H(t)$, it is difficult to determine $V(t)$ and $\tilde V(t)$ such that the effective Hamiltonian $H^V(t)$ takes the diagonal form. We therefore reverse engineer the physical Hamiltonian as
\begin{equation}
H(t)=i\dot{V}(t)V^{-1}(t)-\tilde{V}^{-1}(t)H^{V}(t)V^{-1}(t), \label{H(t)}
\end{equation}
which renders the non-Hermitian dynamics exactly solvable.

In the following, we specialize the general formalism to a three-level system for the implementation of arbitrary single-qubit holonomic gates. Specifically, we introduce the auxiliary frame $\{|\nu_{n}(t)\rangle, |\tilde{\nu}_{n}(t)\rangle\}$ to implement holonomic evolution on a two-dimensional logical subspace, and further specify the complementary pair $\{|\nu^{c}(t)\rangle, |\tilde{\nu}^{c}(t)\rangle\}$ in the remaining one-dimensional subspace. This construction yields the Hamiltonian in Eq. (\ref{H(t)}), which will later be shown to describe a non-Hermitian three-level system with decay and dephasing driven by complex pulses in the conditional no-jump regime. Importantly, unlike in the two-level case, a generic three-level Hamiltonian generally contains a diagonal contribution not proportional to the identity, so that the resulting dynamics is not a trivial extension of the two-level scenario.

\textit{Construction of arbitrary holonomic single-qubit gates}.
We consider a driven three-level system in the ordered bare basis $(|0\rangle,|2\rangle,|1\rangle)$, where $|j\rangle$ has bare energy $E_j$ $(j=0,1,2)$.  In this basis, the three bare states are identified with the canonical basis vectors of $\mathbb{C}^3$. Operators on the three-dimensional Hilbert space can be parameterized by the eight Gell-Mann matrices $\{\lambda_i\}_{i=1}^{8}$ together with the diagonal matrix $E_{3}=\mathrm{diag}(1,1,0)$ \cite{Gell-Mann}. Here we adopt the following factorized ansatz for the transformation operators:
\begin{eqnarray}
V(t)&=&e^{-iA(t)}e^{-i\frac{\alpha(t)}{2}\lambda_{5}}e^{-i\frac{\beta(t)}{2}\lambda_{7}}, \nonumber \\
\tilde{V}(t)&=&e^{i\frac{\beta(t)}{2}\lambda_{7}}e^{i\frac{\alpha(t)}{2}\lambda_{5}}e^{iA(t)},  \label{V}
\end{eqnarray}
Here $A(t)=\frac{1}{2}\Big[(\theta_{1}(t)-\theta_{2}(t))\lambda_{3}+(\theta_{1}(t)+\theta_{2}(t)+\theta_{3}(t))E_{3}-\sqrt{3}\,\theta_{3}(t)\lambda_{8}\Big]$,
with $\theta_{1}(t)=\theta_{3}(t)+\theta$ and $\theta\in\mathbb{R}$. The functions $\theta_{1}(t)$ and $\theta_{2}(t)$ are generally complex and satisfy $\mathrm{Re}[\theta_{2}(t)]=-\mathrm{Re}[\theta_{1}(t)]$, while $\alpha\equiv\alpha_0$ is taken to be a real constant and $\beta(t)$ is a real-valued function.

The logical qubit is encoded in $\mathcal{M}(t_0)=\mathrm{span}\{|0\rangle, |1\rangle\}$, while the auxiliary state $|2\rangle$ spans the complementary subspace $\mathcal{C}(t_0)$.
Using Eq. (\ref{operators}), the corresponding right and dual biorthogonal basis vectors of $\mathcal{M}(t)\bigoplus\mathcal{C}(t)$,
satisfying $\langle\tilde{\nu}_{\mu}(t)|\nu_{\nu}(t)\rangle=\delta_{\mu\nu}$ and $\langle\tilde{\nu}^{c}(t)|\nu^{c}(t)\rangle=1$, are given by
\begin{widetext}
\begin{eqnarray}
\begin{aligned}
|\nu_{0}(t)\rangle&=\left(
\begin{array}{c}
 e^{-i\theta_{1}}\cos\frac{\alpha}{2} \\
 0 \\
 e^{-i\theta_{3}}\sin\frac{\alpha}{2} \\
 \end{array}
 \right), &
|\nu_{1}(t)\rangle&=\left(
\begin{array}{c}
 -e^{-i\theta_{1}}\sin\frac{\alpha}{2}\cos\frac{\beta}{2} \\
 -e^{-i\theta_{2}}\sin\frac{\beta}{2} \\
 e^{-i\theta_{3}}\cos\frac{\alpha}{2}\cos\frac{\beta}{2} \\
 \end{array}
 \right), &
 |\nu^{c}(t)\rangle&=\left(
\begin{array}{c}
 -e^{-i\theta_{1}}\sin\frac{\alpha}{2}\sin\frac{\beta}{2} \\
 e^{-i\theta_{2}}\cos\frac{\beta}{2} \\
 e^{-i\theta_{3}}\cos\frac{\alpha}{2}\sin\frac{\beta}{2} \\
 \end{array}
 \right), & \nonumber
 \end{aligned}
 \\
\begin{aligned}
|\tilde{\nu}_{0}(t)\rangle&=\left(
\begin{array}{c}
 e^{-i\theta^{*}_{1}}\cos\frac{\alpha}{2} \\
 0 \\
 e^{-i\theta^{*}_{3}}\sin\frac{\alpha}{2} \\
 \end{array}
 \right), &
|\tilde{\nu}_{1}(t)\rangle&=\left(
\begin{array}{c}
 -e^{-i\theta^{*}_{1}}\sin\frac{\alpha}{2}\cos\frac{\beta}{2} \\
 -e^{-i\theta^{*}_{2}}\sin\frac{\beta}{2} \\
 e^{-i\theta^{*}_{3}}\cos\frac{\alpha}{2}\cos\frac{\beta}{2} \\
 \end{array}
 \right), &
 |\tilde{\nu}^{c}(t)\rangle&=\left(
\begin{array}{c}
 -e^{-i\theta^{*}_{1}}\sin\frac{\alpha}{2}\sin\frac{\beta}{2} \\
 e^{-i\theta^{*}_{2}}\cos\frac{\beta}{2} \\
 e^{-i\theta^{*}_{3}}\cos\frac{\alpha}{2}\sin\frac{\beta}{2} \\
 \end{array}
 \right). & \label{externals}
 \end{aligned}
\end{eqnarray}
\end{widetext}
In particular, when $\beta(t_0)=0$ one has $|\nu^{c}(t_0)\rangle=e^{-i\theta_2(t_0)}|2\rangle$, whereas $|\nu_{0}(t_0)\rangle$ and $|\nu_{1}(t_0)\rangle$ span
$\{|0\rangle, |1\rangle\}$. It is mentioned that the state $|\nu^{c}(t)\rangle$ does not need the periodic boundary condition.

For a cyclic evolution of subspace satisfying Eq. (\ref{externals}), the induced transformation on the logical subspace takes, up to a physically irrelevant global phase, the form
\begin{equation}
U_{s}(t_f,t_0)=e^{i(\Phi/2)\vec{n}\cdot\vec{\sigma}}, \label{gateU}
\end{equation}
with
\begin{equation}
\Phi=2\int_{t_0}^{t_f} \text{Re}[\dot{\theta}_1(t)]\sin^{2}\frac{\beta(t)}{2}dt. \label{phase1}
\end{equation}
Here $\vec{n}=(\sin\alpha_{0}\cos\theta, \sin\alpha_{0}\sin\theta, \cos\alpha_{0})$ is a unit vector and $\vec{\sigma}=(\sigma_x, \sigma_y, \sigma_z)$ denotes the Pauli matrices. Equation (\ref{gateU}) has the standard $SU(2)$ rotation form. Crucially, the rotation axis $\vec{n}$ is fixed solely by $\alpha_{0}$ and $\theta$, whereas the rotation angle $\Phi$ is determined by the path $L$ in the $\{\beta,\mathrm{Re}[\theta_1]\}$ parameter space. The axis and angle are therefore independently tunable, which establishes universal single-qubit control in the logical subspace. The geometric character of the gate becomes explicit upon rewriting Eq. (\ref{phase1}) as $\Phi=\int_{L} (1-\cos\beta)d\text{Re}[\theta_1]$. For a cyclic evolution, introducing the effective Bloch-sphere coordinates $(\vartheta,\varphi)=(\beta,\text{Re}[\theta_1])$ further gives $\Phi=\int_{L} (1-\cos\vartheta)d\varphi=\Omega$, where $\Omega$ is the solid angle enclosed by $L$, up to the orientation convention. The gate phase is thus geometric in origin, being determined by the enclosed area rather than by the rate at which the path is traversed.

A concrete implementation is provided by a dissipative Rydberg $\Lambda$-type system, where $|0\rangle$ and $|1\rangle$ are two hyperfine ground states and $|2\rangle$ is a Stark eigenstate. Conditioned on no quantum jump, the effective Hamiltonian within the rotating-wave approximation reads
\begin{equation}
\begin{split}
H(t)&=-i\Gamma_2(t)|2\rangle\langle2|+\sum_{j=0}^{1}\{-[\Delta_{j}(t)+i\Gamma_{j}(t)]|j\rangle\langle j|\\
&+\tilde{\Omega}_{j}^{+}(t)|j\rangle\langle2|+\tilde{\Omega}_{j}^{-}(t)|2\rangle\langle j|\}, \\
\end{split}
\label{eq:rydberg_H}
\end{equation}
where the effective complex Rabi amplitudes are defined as
\begin{equation}
\tilde{\Omega}_j^{\pm}(t)=\Omega_j^0(t)e^{\pm i\phi_j^0(t)}+ i\,\Omega_j^1(t)e^{\pm i\phi_j^1(t)}.\label{eq:complex_rabi}
\end{equation}
Under the two-photon-resonance condition, $\Delta_0(t)=\Delta_1(t)\equiv\Delta(t)$. Matching Eq. (\ref{H(t)}) yields
\begin{eqnarray}
\tilde{\Omega}_0^{\pm}(t)&=&-\frac{1}{2}\sin\frac{\alpha}{2}\ \chi_{\pm}(t)e^{\pm i[\theta_2(t)-\theta_1(t)]}, \nonumber \\
\tilde{\Omega}_1^{\pm}(t)&=&\frac{1}{2}\cos\frac{\alpha}{2}\ \chi_{\pm}(t)e^{\pm i[\theta_2(t)-\theta_3(t)]},  \label{condition} \\
\Delta(t)&=&f_2(t)\sin^2\frac{\beta(t)}{2}+f_3(t)\cos^2\frac{\beta(t)}{2}-\mathrm{Re}\,\dot{\theta}_1(t), \nonumber
\end{eqnarray}
with $\chi_{\pm}(t)=[f_{3}(t)-f_{2}(t)]\sin\beta(t)\pm i\dot{\beta}(t)$. Most importantly, within the no-jump dynamics the intrinsic dissipation is matched exactly by the control through
\begin{equation}
\Gamma_{i}(t)=-\mathrm{Im}\dot{\theta}_{i}(t),
\label{dissipation}
\end{equation}
so that the physical decay and dephasing rates enter the pulse design as known inputs rather than as uncontrolled perturbations.
Since the complex envelopes $\widetilde{\Omega}_j^{\pm}(t)$ and the detuning $\Delta(t)$ remain externally programmable, a suitable choice of $\{\theta_{1},\theta_{2},\theta_{3},f_{2},f_{3},\alpha_0,\beta\}$ makes the driven $\Lambda$ system follow exactly the target non-Hermitian trajectory constructed above.

Accordingly, although the conditional evolution in the full three-level Hilbert space is generally nonunitary, the encoded logical subspace still implements exactly the target unitary gate within the no-jump sector,
\begin{eqnarray}
\begin{aligned}
U_{\mathcal{M}}(t_f,t_0)&=e^{i(\Phi/2)\vec{n}\cdot\vec{\sigma}}, &
U_{\mathcal{M}}^{\dagger}(t_f,t_0)U_{\mathcal{M}}(t_f,t_0)&=I_2. \label{eq:logical_gate}
\end{aligned}
\end{eqnarray}
Although phrased here for a Rydberg atom, the construction applies more broadly to driven $\Lambda$ systems with independently programmable complex couplings and monitored dissipation.

To make the mechanism concrete, we consider the NOT gate as a representative example. We choose $\alpha\equiv\pi/2$, $\beta(t)=2\arcsin[(1/{\sqrt{2}})\sin({\xi t}/2)]$, $\text{Re}[\theta_{1}(t)]=\xi t$, and $\theta=0$, with $\xi t\in[0,2\pi]$. These choices give the geometric phase $\Phi=\pi$ and the rotation axis $\vec{n}=(1, 0, 0)$, so that Eq. (\ref{gateU}) yields $U_{s}=i\sigma_{x}$. Apart from the overall phase factor $e^{i\pi/2}$, this operation is physically equivalent to the NOT gate $\sigma_x$, namely, $|a\rangle\mapsto|a\oplus1\rangle$. For the short gate duration considered here,
$\Delta t=2\pi/\xi=100\,\mathrm{ns}$, the intrinsic dissipative rates vary negligibly and can therefore be treated as constants. We take $\Gamma_{2}=4\Gamma_{1}=4\Gamma_{3}=4\pi\,\mathrm{kHz}$ \cite{rate1,rate2}.  A corresponding choice of the remaining parameters is as follows: $\text{Im}[\theta_{2}(t)]=- 4\Gamma_{1}t$ for $\xi t\in[0,2\pi]$, while $\text{Im}[\theta_{1}(t)]=- \Gamma_{1}t$ for $\xi t\in[0,\pi]$ and $\text{Im}[\theta_{1}(t)]=(2\pi-\xi t)\Gamma_{1}/\xi$ for $\xi t\in[\pi,2\pi]$, with $\Gamma_{1}/\xi=5\times10^{-5}$. The nonuniqueness of the auxiliary functions $f_{2,3}(t)$ provides a residual freedom that will be exploited in the next section to optimize the robustness of the gate against Rabi-amplitude and detuning errors.

\textit{Robustness Against Systematic Errors and Quantum Jumps}.
We first examine an analytically tractable class of systematic errors in the conditioned no-jump dynamics, namely quasi-static phase miscalibrations in $\theta_{i}(t)$ $(i=1,2,3)$. Since these fluctuations are much slower than the gate-operation time, we model them as quasi-static real offsets, $\theta^{'}_{i}(t)=\theta_{i}(t)+\delta\theta_{i}$ with $|\delta\theta_{i}|\ll1$ and $\delta\dot{\theta}_{i}=0$. Substituting $\theta^{'}_{i}(t)$ into Eq. (\ref{condition}) shows that the detuning $\Delta(t)$, the magnitudes of the effective couplings $\tilde{\Omega}_i^{\prime\,\pm}(t)$, and the holonomic angle $\Phi$ defined in Eq. (\ref{phase1}) remain unchanged; the only effect is to imprint static phase shifts on the driving fields. In terms of the physical pulse phases, this corresponds to $\delta\phi_0^0=-\delta\phi_0^1=\delta_{21}$ and $\delta\phi_1^1=-\delta\phi_1^0=\delta_{32}$, where $\delta_{21}=\delta\theta_{2}-\delta\theta_{1}$ and $\delta_{32}=\delta\theta_{3}-\delta\theta_{2}$. A deviation in $\delta\theta_{2}$ alone is therefore gauge-like: it changes the optical phases of the pump and Stokes pulses but leaves the logical rotation axis unchanged. By contrast, the physically relevant phase error $\delta\theta$ tilts the rotation axis from $\bm{n}(\alpha_{0},\theta)$ to $\bm{n}'(\alpha_{0},\theta+\delta\theta)$ and thereby reduces the gate fidelity.

To quantify this effect, we use the normalized trace fidelity $F_{s}=|\text{Tr}(V_{s}U^{\dagger}_{s})|/|\text{Tr}(V_{s}V^{\dagger}_{s})|$ with $V_{s}=e^{-i(\Phi/2)\bm{n}'\cdot\bm{\sigma}}$. Here $U_{s}$ is the ideal single-qubit holonomy. Expanding to second order in $\delta\theta$, we obtain $F_{s}\approx1-(1/2)\sin^{2}(\Phi/2)\sin^{2}\alpha_0(\delta\theta)^{2}$. Thus, for the same phase-miscalibration model, the second-order infidelity is no larger than that of the corresponding dynamical gate $F_{d}=1-(1/2)\sin^{2}\alpha_{0}(\delta\theta)^{2}$, or the conventional holonomic gate $F_{g}=1-(1/2)\sin^{2}(\Phi/2)(\delta\theta)^{2}$ as discussed in Refs. \cite{dynamical,holonomic1}. The physical origin of this enhanced robustness is transparent: the quasi-static phase error does not renormalize the holonomic angle, but only tilts the rotation axis. The resulting suppression factor $\sin^{2}(\Phi/2)\sin^{2}\alpha_0$ shows that the present non-Hermitian nonadiabatic holonomic gate preserves, and can even improve upon, the intrinsic error resilience of geometric control despite arising from conditional non-Hermitian dynamics.

In addition to quasi-static phase offsets, we consider systematic errors in the Rabi amplitudes and detuning. Because the off-diagonal elements of Eq. (\ref{eq:rydberg_H}) are determined by the effective Rabi couplings, amplitude miscalibration is modeled as $\tilde{\Omega}_i^{\pm}(t)\rightarrow(1+\delta_{r})\tilde{\Omega}_i^{\pm}(t)$ for $i=0,1$, while detuning errors are incorporated as $\Delta(t)\rightarrow\Delta(t)+\delta_{d}\Omega_{\text{max}}$, with $\Omega_{\text{max}}$ the maximum instantaneous coupling strength of the ideal protocol. Let $U$ and $V$ denote the ideal and perturbed no-jump propagators over one gate cycle. To quantify robustness in the dissipative setting, we evaluate the normalized fidelity $F_{e}(\eta)=|\langle\psi(\eta)|VU^{\dagger}|\psi(\eta)\rangle|/|\langle\psi(\eta)|VV^{\dagger}|\psi(\eta)\rangle|$, and average it uniformly over the logical input states $\psi(\eta)=\cos\eta|0\rangle+\sin\eta|1\rangle$, with $\eta\in[0,\pi/2]$.

As a representative example, we optimize the NOT gate by introducing into the control functions $f_{1,2,3}(t)$ a four-parameter modulation
\begin{eqnarray}
\begin{aligned}
A(t)&=1+\sum_{j=1}^{4} 2j a_j\cos(2j\lambda(t)), &
\lambda(t)&=\frac{\xi t-\sin\xi t}{2}, \label{parameter}
\end{aligned}
\end{eqnarray}
where $a_j$ are variational coefficients. The resulting parametrization, $f_{1}(t)=-\xi[1+2A(t)\cos^{2}(\xi t/2)]$, $f_{2}(t)=\xi[-1+A(t)\sin^{2}(\xi t/2)]$ and $f_{3}(t)=-\xi[1+A(t)(1+\cos^{2}(\xi t/2))]$ is constructed so that $\Phi=\int_{t_0}^{t_f}(f_1(t)-f_3(t))dt=\pi$, thereby fixing the target holonomic angle. This constrained deformation preserves the geometric character of the gate while providing sufficient freedom to reduce sensitivity to both amplitude and detuning errors. We optimize $\{a_j\}$ against the worst-case averaged fidelity at $(\delta_{r},\delta_{d})=(\pm 0.1,0)$ and $(0,\pm 0.1)$. One set of optimized parameters obtained is $a_1=-0.155613$, $a_2=-0.057894$, $a_3=-0.027937$ and $a_4=-0.008838$, for which $\bar{F_{e}}>0.99$ in all four cases. This demonstrates robustness against independently applied $10\%$ amplitude or detuning errors.

A more stringent test is obtained by including genuine quantum jumps. In the Markovian regime, the full open-system dynamics is governed by the Lindblad master equation, whereas the corresponding conditional no-jump evolution is generated by an effective non-Hermitian Hamiltonian \cite{Lindblad,Dalibard}. For conventional real drives, the anti-Hermitian part distorts the target trajectory and reduces the gate fidelity. Here, by contrast, the time-dependent complex drive is designed to cancel this distortion in the conditional evolution, so that residual infidelity arises only from stochastic jump events. The density operator therefore obeys
\begin{equation}
\dot{\rho}(t)=-i(H_{\mathrm{eff}}(t){\rho}-{\rho}H_{\mathrm{eff}}^{\dag}(t))+\sum_{i=0}^{1}(L^{c}_{i2}{\rho}L^{c\dag}_{i2}+L^{p}_{i2}{\rho}L^{p\dag}_{i2}), \label{Liouville1}
\end{equation}
where $H_{\mathrm{eff}}(t)\equiv H(t)$ is the effective non-Hermitian Hamiltonian introduced above. $L^{c}_{i2}=\sqrt{\gamma_{i}}|i\rangle\langle2|$ and $L^{p}_{i2}=\sqrt{\epsilon_{i}}(|2\rangle\langle2|-|i\rangle\langle i|)$ describe decay and dephasing, respectively. To quantify the effect of jumps, we introduce the fidelity
\begin{equation}
\bar{F}_J(t)=\frac{\sqrt{|\langle\psi|U^{\dag}(t)\rho(t)U(t)|\psi\rangle|}}{\sqrt{|\text{Tr}\rho(t)\langle\psi|U^{\dag}(t)U(t)|\psi\rangle|}}, \label{fidelity 1}
\end{equation}
which is averaged uniformly over the logical input manifold $\psi(\eta)=\cos\eta|0\rangle+\sin\eta|1\rangle$, with $\eta\in[0,\pi/2]$.

For the NOT gate, we take representative kHz-scale decoherence rates $\epsilon_{0}=\epsilon_{1}=\gamma_{0}=\gamma_{1}=2\pi\times1\,\mathrm{kHz}$ \cite{rate1,rate2}, and the scanning frequency is $\xi=2\pi\times10\,\mathrm{MHz}$, corresponding to a gate time $\Delta t=2\pi/\xi=100\,\mathrm{ns}$. Numerical integration of Eq. (\ref{Liouville1}) yields an averaged final fidelity $\bar{F}_{J}(t_f)=99.98\%$. Thus, for the representative parameters considered here, compensating the anti-Hermitian part of the conditional no-jump evolution renders the stochastic-jump contribution the dominant residual source of gate error, while this contribution remains small on the gate timescale. This suggests that the protocol retains good robustness beyond the no-jump approximation within the parameter regime considered here.

\section{Conclusion} In summary, we have developed a general theoretical framework for non-Hermitian nonadiabatic holonomic quantum computation and applied it to a non-Hermitian three-level system, in which arbitrary single-qubit logical gates can be implemented. Within a biorthogonal framework, tailored complex control pulses enforce exact closure of the computational-subspace evolution at the target time despite the intrinsically nonunitary no-jump dynamics, thereby enabling nonadiabatic holonomic gates without requiring cyclic evolution in the orthogonal complement. Moreover, by incorporating decay and dephasing of both the computational and auxiliary states directly into the pulse design, our scheme ensures that dissipation does not reduce the no-jump gate fidelity. Combined with a physically motivated Rydberg-atom implementation, these results demonstrate that dissipation-control matching can simultaneously preserve the geometric nature of the gates and provide robustness against representative systematic errors and full quantum-jump effects. Our work extends holonomic quantum control to a genuinely dissipative non-Hermitian setting and opens a practical route toward fast and resilient geometric quantum information processing in realistic open quantum systems.
\bigskip

\acknowledgments
This work is supported by the National Natural Science Foundation of China (Grant No. 12364060) and the Scientific Research Program of Education Department of Shaanxi Provincial Government (Grant No. 22JK0617).

\end{document}